\documentstyle[epsfig,amssymb,twocolumn]{mn2e}
\newif\ifAMStwofonts

\newcommand{\vv}{\mbox{\boldmath$v$}}
\newcommand{\BB}{\mbox{\boldmath$B$}}
\newcommand{\bb}{\mbox{\boldmath$b$}}

\title[Magnetic fields and the rotation curve]
{Magnetic fields: Impact on the rotation curve of the Galaxy}
\author[S\'anchez-Salcedo and Santill\'an]
{F.J. S\'{a}nchez-Salcedo$^{1}$ and A. Santill\'an$^{2}$\\
$^{1}$Instituto de Astronom\'\i a, Universidad Nacional
Aut\'onoma de M\'exico,\\ Ciudad Universitaria, Aptdo. 70 264,
C.P. 04510, Mexico City, Mexico; jsanchez@astro.unam.mx\\
$^{2}$Direcci\'on General de C\'omputo y  Tecnolog\'{\i}as de 
la Informaci\'on y Comunicaci\'on, \\Universidad Nacional
Aut\'onoma de M\'exico, Ciudad Universitaria, C.P.~04510,
 Mexico City, Mexico; alfredo@astro.unam.mx}

\begin{document}

\date{Accepted xxxx Month xx. Received xxxx Month xx; in original form
2009 December 10}
\pagerange{\pageref{firstpage}--\pageref{lastpage}} \pubyear{2009}
\maketitle

\label{firstpage}

\begin{abstract}
We quantify the effects of magnetic fields, cosmic rays
and gas pressure on the rotational velocity
of H\,{\sc i} gas in the Milky Way, at galactic distances
between $R_{\odot}$ and $2R_{\odot}$. 
The magnetic field is modelled by two
components; a mainly azimuthal magnetic component
and a small-scale tangled field. We construct a range of
plausible axisymmetric models consistent with the strength of
the total magnetic field as inferred from radio synchrotron data. 
In a realistic Galactic disc, 
the pressure by turbulent motions, cosmic rays and the tangled turbulent
field provide radial support to the disc.  
Large-scale (ordered) magnetic fields may or may not provide
support to the disc, depending on the local radial gradient
of the azimuthal field.  We show that for observationally
constrained models, magnetic forces cannot appreciably alter the 
tangential velocity of H\,{\sc i} gas 
within a galactic distance of $2R_{\odot}$.

\end{abstract}

\section{Introduction}

The interstellar medium in galaxies contains three basic constituents:
ordinary matter, cosmic rays and magnetic fields. Studies of the vertical
distribution of gas and synchrotron emission in the solar neighbourhood
show that cosmic rays and magnetic fields influence the spatial
distribution of gas providing efficient support
against the gravitational force (e.g., Ferri\`ere 2001; Cox 2005).
In the radial direction, gradients in the pressure may produce
a difference between the rotational velocity of the gas $v_{\phi}$ and the
real gravitational circular velocity $v_{c}$,
defined as $v_{c}^{2}\equiv Rd\Phi/dR$.
Here $R$ is the galactocentric radius and $\Phi$
the gravitational potential.
The asymmetric drift, defined as $v_{\phi}^{2}-v_{c}^{2}$,
measures this difference.
In a gaseous disc in equilibrium, the asymmetric drift is a 
consequence of the support
by thermal, turbulent and magnetic pressures as well as the
pressure due to cosmic rays (e.g., Parker 1966; Spitzer 1978).
In galaxies with circular velocities $v_{\phi}>50$ km s$^{-1}$, 
the asymmetric drift corrections to derive the real
gravitational circular velocity from the observed rotational
velocity are not applied because 
they are small as compared to uncertainties due to inclination,
warps, non-circular motions, etc (e.g., de Blok \& Bosma 2002).
Only for low-mass galaxies with $v_{\phi}<50$ km s$^{-1}$,
corrections for the asymmetric drift must be taken into account
(e.g.~Dalcanton \& Stilp 2010).

In this approach, magnetic effects on the gas 
are modelled as a pressure term in the asymmetric drift.
However, gas can also experience an additional force due to the magnetic
stress of a large-scale magnetic field.
Using a stationary cylindrical model, Nelson (1988) argued that the
dynamical effects of magnetic fields can be very significant, yielding
rotational velocities significantly higher than the gravitational
orbital velocity, because of the inward force due to the magnetic
tension. His model, however, predicted unrealistic radial velocities of
the gas ($\sim 200$ km s$^{-1}$ at $R \sim 30$ kpc), 
because of the magnetic torque.
Assuming a purely azimuthal magnetic field,
Battaner et al.~(1992) 
derived the magnetic field strength as a function of galactocentric
radius required to explain the rotation curve of M31 without any dark matter.
Still, the field needed is so strong that the
magnetic pressure in the vertical direction 
would cause the gaseous disc to flare unacceptably
(Cuddeford \& Binney 1993) and, thus, magnetic fields
are not a real alternative to dark matter. 

In a more conventional scenario, S\'anchez-Salcedo (1997a)
combined the effects of an azimuthal magnetic field of strength 
$\sim 1 \mu$G with an isothermal dark halo to fit reasonably
well the detailed shape of the rotation curve of the dwarf
galaxy NGC 1560. 
By constructing models that match boundary conditions
at infinity, S\'anchez-Salcedo \& Reyes-Ruiz (2004)
found that the magnetic contribution cannot boost the azimuthal
speed of the gas by more than $\sim 20$ km s$^{-1}$ at the outermost point
of H\,{\sc i} detection.

The idea that galactic magnetic fields can alter 
the rotation curves of spirals has been
revived recently.
Beck (2007) suggests that the 
low decrease of the magnetic field energy density in
the galaxy NGC 6946 to large radii may affect the gas dynamics 
in the outer galaxy.
Recently, Ruiz-Granados et al.~(2010, 2012) claim 
that large-scale magnetic fields can
provide enough radial confinement of the gas to
explain the rising-up in the H\,{\sc i} rotation curve 
detected in some galaxies.
Moreover, they argue that the shape of the H\,{\sc i} rotation curves 
of M31 and 
the Milky Way are fitted better if the contribution of 
the large-scale (mainly azimuthal) magnetic field is included.
Tsiklauri (2011) uses a bisymmetric spiral configuration
to model the magnetic field of the Milky Way and concludes that
the magnetic pinching effect may be important for $R\geq 15$ kpc. 
Ja{\l}ocha et al.~(2012a,b) suggest that the mass-to-light ratio
in the discs of the galaxies NGC 891 and NGC 253 are more realistic 
if the contribution of magnetic fields give rise to a faster
circular velocity.

It remains unclear how these findings are compatible with rigorous
upper limits based on the Virial Theorem arguments indicating that
magnetic fields can hardly speed up H\,{\sc i} discs by more than
$20$ km s$^{-1}$ in the outermost point of H\,{\sc i} detection
(S\'anchez-Salcedo \& Reyes-Ruiz 2004). 
Since the relative importance of turbulent, magnetic and
cosmic ray pressures is comparable,
it is certainly not clear which is the role of the pressure 
by cosmic rays and the tangled component of the magnetic field
in providing support to the disc.
In this paper, we combine different observations to 
explore whether and how magnetic fields and cosmic rays
can alter the gas rotation curve in the Milky Way.

The paper is structured as follows. In Section \ref{sec:assump}, we describe
the formalism, our simplifying assumptions and the basic equations. In Section
\ref{sec:models}, we highlight the reasons why the Milky Way is an 
excellent target to carry out this analysis and present a range of plausible
magnetic models compatible with observations. In Section \ref{sec:results} 
we calculate the expected differences between the observed rotation curve
and the true gravitational circular velocity for these models.
Conclusions are given in Section \ref{sec:conclusions}.

\section{Assumptions and governing equations}
\label{sec:assump}
We consider a magnetized disc of slightly ionised gas with the axis along
the $z$--direction that is described well in the ideal magnetohydrodynamic 
limit (MHD). 
In galactic discs, the cosmic ray population forms a light fluid 
with significant 
pressure, which is coupled via magnetic fields to the thermal
interstellar components. Hence, the pressure by cosmic rays 
may help to support thermal gas and will be therefore included.

Following previous works, we will assume that the disc is axisymmetric
over time (S\'anchez-Salcedo 1997a; Ja{\l}ocha et al.~2012;
Ruiz-Granados et al.~2010, 2012).
Non-axisymmetric configurations
are more difficult to deal with because they generate magnetic density
waves (e.g., Lou \& Fan 1998).
Axisymmetry is the simplest assumption to quantify
the overall effect of magnetic fields on the azimuthally averaged tangential
velocity $v_{\phi}$ in an equilibrium configuration. 

It is assumed that the magnetic field 
can be decomposed into an average part $\bar{\BB}(R)$ varying only
on the large scale and a small-scale isotropic random field $\bb$,
so that $\left<\bb\right>=0$. We will refer to $\left<b^{2}\right>^{1/2}$
as the strength of the random (or turbulent) magnetic field. 
At scales larger than the coherence length of the small-scale magnetic
field, it is useful to define the strength of the total 
magnetic field as $B_{\rm tot}^{2}=\bar{B}^{2}+\left<b^{2}\right>$. 
In the equilibrium configuration, we assume that the regular magnetic field 
consists of a planar magnetic field 
$\bar{\BB}(R)=(\bar{B}_{R},\bar{B}_{\phi},0)$ 
(in cylindrical coordinates), with $\bar{B}_{R}\ll \bar{B}_{\phi}$,
We further assume that the radial velocity of the gas, $v_{R}$, is
much smaller than $v_{\phi}$, and thus it can be ignored;  
so the velocity in the disc is $\vv=(0,v_{\phi},0)$ in 
cylindrical coordinates\footnote{This is an approximation because
the magnetic field creates a torque, unless $\bar{B}_{R}=0$,
leading to a radial inflow of gas
(S\'anchez-Salcedo 1997b).}.
In principle, each component of the interstellar gas can rotate at
different velocity. Since we are only interested in the rotation curve
of neutral atomic gas, we will consider the dynamics
of this component and ignore the presence of molecular hydrogen gas.  
In the Milky Way, this is a good approximation especially at $R>10$ kpc
because it is at these galactic distances where the neutral atomic hydrogen is 
dominant in the mass budget of the interstellar gas.

Because of the symmetry around $z=0$, we take that all the derivatives 
with respect to $z$ are negligible near the midplane of the disc. 
Under these circumstances, the radial component of the 
motion equation of the gas at $z=0$ reads
\begin{equation}
v_{\phi}^{2}=v_{c}^{2}+v_{P}^{2}+
v_{\rm mag}^{2},
\end{equation}
where $v_{\rm mag}^{2}$ is the contribution of the
regular (azimuthal) magnetic field: 
\begin{equation}
v_{\rm mag}^{2}\equiv
\frac{1}{8\pi R \rho} \frac{d(R^{2}\bar{B}_{\phi}^{2})}{dR}
=\frac{R}{4\pi \rho}\left(\frac{\bar{B}_{\phi}^{2}}{R}+
\bar{B}_{\phi}\frac{d\bar{B}_{\phi}} {dR}\right),
\label{eq:vmag}
\end{equation} 
and $v_{P}^{2}$ is the contribution by pressure gradients, 
\begin{equation}
v_{P}^{2}\equiv \frac{R}{\rho}\frac{dP_{T}}{dR},
\label{eq:vP}
\end{equation}
where $\rho$ is the gas volume density at the midplane and 
$P_{T}(R)=P_{g}+P_{b}+P_{\mathrm CR}$ 
is the total gas pressure consisting of the gas kinetic
pressure (thermal plus turbulent), 
the magnetic pressure $P_{b}$, arising from the random magnetic
field component plus also the pressure by cosmic rays $P_{CR}$
(the pressure by radiation will be ignored). 
More specifically,
the kinetic pressure $P_{g}$ is given by 
\begin{equation}
P_{g}=\rho\sigma^{2},
\label{eq:kin_pressure}
\end{equation}
where $\sigma$ is the H\,{\sc i} line width in the 
radial direction, which is approximately constant
or slightly decreasing with $R$ in the outer parts of the H\,{\sc i} discs, 
typically $\sigma\simeq 6-8$ km s$^{-1}$ (e.g., Dib et al.~2006, and 
references therein; see Blitz \& Spergel 1991 and Burton 1992, 
for our Galaxy). 
The spatially averaged magnetic pressure by the turbulent field is taken as
\begin{equation}
P_{b}=\frac{\left<b^{2}\right>}{8\pi}.
\end{equation}
Finally, the cosmic-ray pressure is expected to be proportional to
magnetic pressure:
\begin{equation}
P_{CR}=\frac{\mu B_{\rm tot}^{2}}{8\pi},
\label{eq:cr_pressure}
\end{equation} 
where $B_{\rm tot}$ is the strength of the total magnetic field 
(that is, $B_{\rm tot}^{2}=\bar{B}^{2}+\left<b^{2}\right>$)
and $\mu$ is a constant of the order of $1$.
This is justified by minimum-energy-type arguments (e.g., Beck et al.~1996).

We want to stress that $v_{P}^{2}$ and $v_{\rm mag}^{2}$ are not necessarily
positive quantities. For instance, an unmagnetized isothermal disc with a 
radially decreasing density has $v_{P}^{2}<0$, which signifies that it provides
pressure support to the disc (i.e.~$v_{\phi}<v_{c}$) because it produces
a force pointing outward.
From Equation (\ref{eq:vmag}), it is simple to see that
the magnetic tension imposes an inward force, 
i.e.~$v_{\rm mag}^{2}>0$,
provided that the azimuthal magnetic field decays radially not
faster than $1/R$.  
Note that in the axisymmetric case with $\bar{B}_{z}=0$, the divergence-free
condition implies $\bar{B}_{R}\propto 1/R$. Therefore, if the 
magnetic pitch angle is constant with $R$, we infer $\bar{B}_{\phi}\propto 1/R$,
implying that $v_{\rm mag}^{2}=0$.
Consequently, a radial decay of $\bar{B}_{\phi}$ slower than
$1/R$ requires a pitch angle decreasing with $R$.

To study the distribution of mass of a certain galaxy, we need $v_{c}$,
which is the circular speed of a test particle, but what we observe
is the azimuthal velocity of the gas $v_{\phi}$. 
For which values of $v_{P}^{2}+v_{\rm mag}^2$ is the correction
to the rotation curve significant?
As guide numbers, if we observe that the gas rotates at
a given galactocentric radius with
a tangential velocity of $v_{\phi}=230$ km s$^{-1}$
and $v_{P}^{2}+v_{\rm mag}^2\simeq 9000$
km$^{2}$ s$^{-2}$, then the gravitational
circular velocity is $v_{c}\simeq 210$ km s$^{-1}$.   
Hence, values of $9000$ km$^{2}$s$^{-2}$ produce a boost
of $20$ km s$^{-1}$.
On the other hand, if $v_{P}^{2}+v_{\rm mag}^2$ was $-9000$
km$^{2}$ s$^{-2}$, then $v_{c}=249$ km s$^{-1}$.
For a low-mass galaxy with $v_{\phi}=120$ km s$^{-1}$, a value of 
$v_{P}^{2}+v_{\rm mag}^2\simeq 4400$ km$^{2}$ s$^{-2}$ implies
that $v_{c}=100$ km s$^{-1}$.
It is $v_{P}^{2}+v_{\rm mag}^2$ that we want to calculate in the 
Milky Way.

\section{The Milky Way as a test case: Models}
\label{sec:models}
The determination of the distribution of H\,{\sc i} volume density and the
magnetic structure of our Galaxy has been improved significantly
over the last two decades. Therefore, our Galaxy is a natural laboratory
to quantify the effects of magnetic fields in the gas dynamics (see
also Vall\'ee 1994). In order to estimate $v_{P}^{2}$ and
$v_{\rm mag}^{2}$ we need
to know the azimuthally averaged radial distribution of the
H\,{\sc i} volume density at the midplane, and the radial profile
of both $\bar{B}_{\phi}$ and $\left<b^{2}\right>$.

\begin{figure}
\epsfig{file=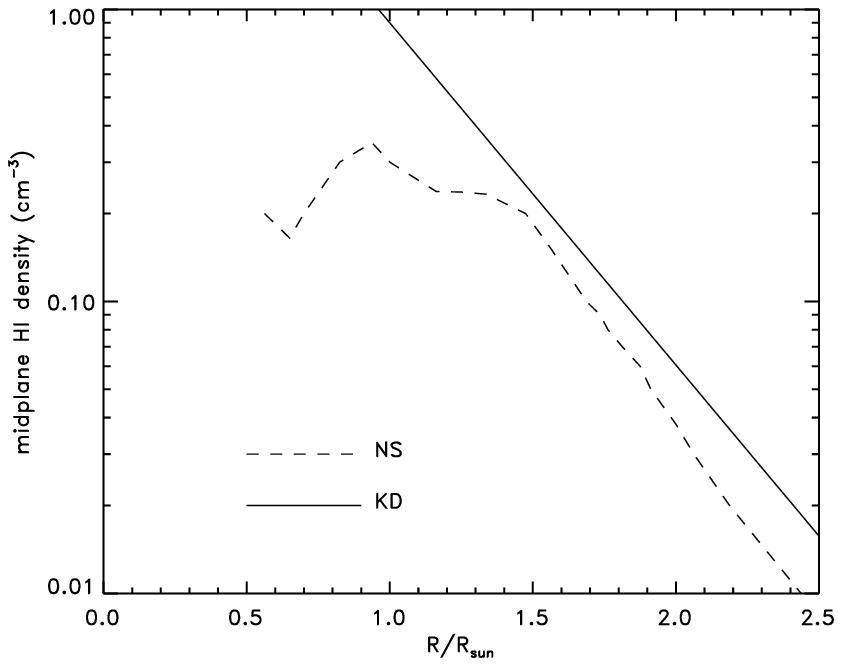,angle=0,width=8cm,height=7cm}
\caption{Midplane H\,{\sc i} volume density of the Milky Way taken
from KD (solid line) and NS (dashed line). 
The KD curve corresponds to an exponential fit to the data.}
  \label{fig:HIdensities}
\end{figure}

The azimuthally averaged H\,{\sc i} volume density 
at the midplane has been derived by Nakanishi \& Sofue (2003, hereafter NS) 
using the Leiden/Dwingeloo survey, the Parkes survey and the NRAO survey.
More recently, Kalberla \& Dedes (2008, hereafter KD) 
inferred the average H\,{\sc i} density at the midplane excluding extra-planar
gas, using the Leiden/Argentine/Bonn (LAB) H\,{\sc i} line survey,
which combines the southern sky survey of the Instituto Argentino
de Radioastronom\'{\i}a (IAR) with an improved version of the
Leiden/Dwingeloo survey. 
NS and KD used different assumptions to derive the H\,{\sc i} 
structure of the Milky Way.
NS chose as Galactic constants 
$R_{\odot}=8$ kpc and $v_{\odot}=217$ km s$^{-1}$, and adopted 
a slightly declining rotation curve to convert the observed brightness
temperature distribution to density, whereas
KD used $R_{\odot}=8.5$ kpc and $v_{\odot}=220$ km s$^{-1}$
and an almost flat rotation curve.  
In addition, NS assumed cylindrical rotation along $z$, whereas
KD adopted a ``lagging'' halo. A discussion about the impact of the different 
assumptions on the H\,{\sc i} distribution can be found in Kalberla et 
al.~(2007). 

\begin{figure*}
\epsfig{file=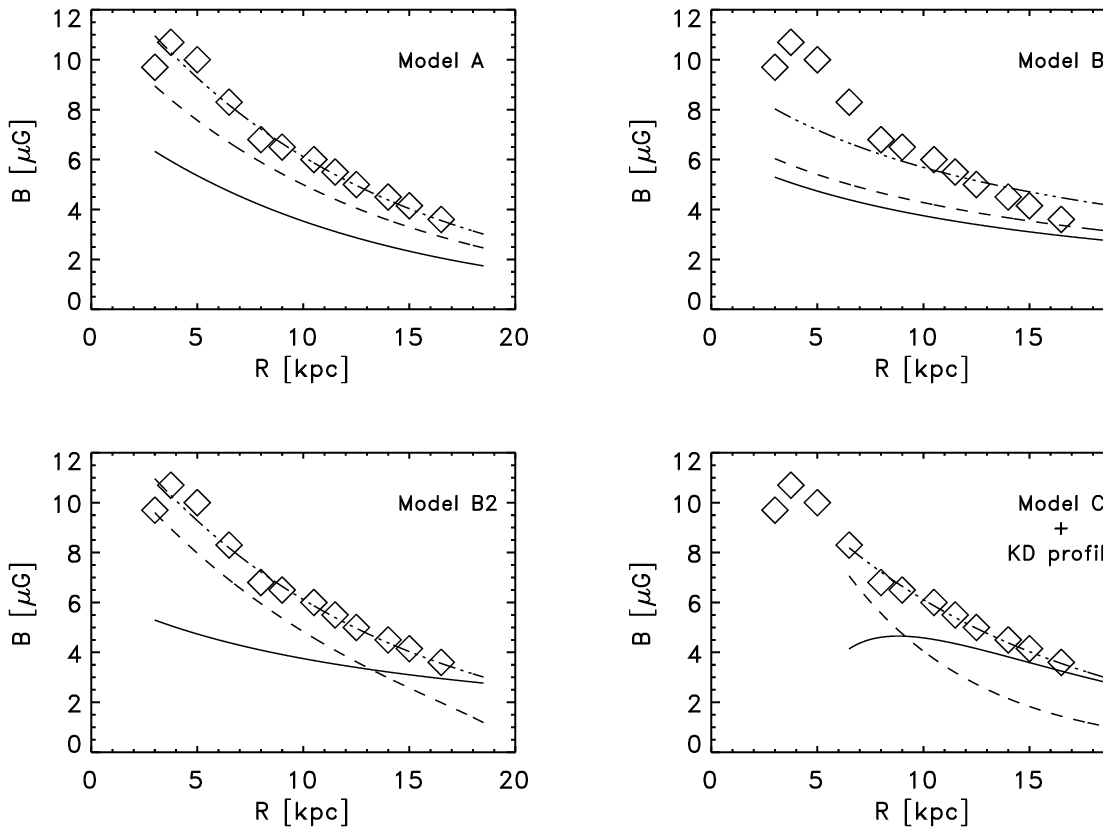,angle=0,width=17cm,height=14cm}
  \caption{Variation of the total ($B_{\rm tot}$; dot-dashed lines), 
regular ($\bar{B}_{\phi}$; solid lines)
and turbulent ($\left<b^{2}\right>^{1/2}$; dashed lines) magnetic 
field strengths for the
different models. In this plot, we have used $R_{\odot}=8.5$ kpc and
the reference KD model. 
The symbols show the strength
of the total magnetic field from the deconvolved surface brightness
of synchrotron emission at $408$ MHz (see text). }
  \label{fig:B_profiles}
\end{figure*}

In Figure \ref{fig:HIdensities} we plot the midplane
H\,{\sc i} density distributions versus $R/R_{\odot}$, as
derived in NS. We see that the density decays exponentially beyond
$1.5R_{\odot}$. On the other hand,
KD found that, for $7\leq R\leq 35$ kpc, the midplane H\,{\sc i}
density can be approximated by 
$n_{H}=n_{\odot}\exp[-(R-R_{\odot})/R_{H}]$ with $n_{\odot}=0.9$ cm$^{-3}$
and $R_{H}=3.15$ kpc.  For comparison, 
we also plot the exponential fit as derived by KD.  
in Figure \ref{fig:HIdensities}. 
We see that the major discrepancies between NS and KD occur 
inside $1.5R_{\odot}$. KD derived an H\,{\sc i} plateau in surface
density of $10M_{\odot}$pc$^{-2}$ at the inner Galaxy, which fits
better to what is known for external galaxies than the saturation value
derived in NS, of $2M_{\odot}$pc$^{-2}$. 
Therefore, we will use KD as our reference H\,{\sc i} gas model but also
refer to NS to explore the effect of systematic uncertainties in the
the derivation of the midplane H\,{\sc i} density.
In order to include 28\% of helium and 1.5\% of
heavier elements, we will convert $n_{H}$ in mass density using
the relation $\rho=1.4m_{p}n_{H}$,
where $m_{p}$ is the proton rest mass (e.g., Ferri\`ere 2001; Cox 2005).

The magnetic field of the Galaxy has been studied through synchrotron emission,
Faraday rotation, optical polarization and Zeeman splitting. The strength
of the total magnetic field averaged in azimuth, $B_{\rm tot}$, 
was obtained from the
surface brightness of synchrotron emission at 408 MHz. In the radial interval
between $3.5$ kpc and $17$ kpc (using $R_{\odot}=8.5$ kpc), and assuming energy
equipartition between magnetic fields and cosmic rays, the
strength of the total magnetic field in the disc can be fitted by
\begin{equation}
B_{\rm tot,fit}=B_{\rm tot,\odot}\exp\left(-\frac{R-R_{\odot}}{R_{B}}\right),
\label{eq:tot_fit}
\end{equation}
where $B_{\rm tot,\odot}$ is the total magnetic field near the Sun,
which is about $6\pm 2$ $\mu$G, and $R_{B}=12$ kpc
(Beck et al.~1996; Strong et al.~2000;
Beck 2001; Ferri\`ere 2001; see also Jansson \& Farrar 2012 using
WMAP7 22 GHz data). 
Since there is no
reliable observational measurement of the magnetic field strength
beyond $2R_{\odot}$,
we will restrict our analysis to $R<2R_{\odot}$, the same
interval studied in Ruiz-Granados et al. (2012).

In order to estimate $v_{\rm mag}^{2}$ and $P_{b}$ we need to separate
the ordered magnetic field and the turbulent magnetic field components.
Starlight and synchrotron polarization data suggest
that the local ratio between the regular and the total magnetic fields is  
$0.6-0.7$ (e.g., Berkhuijsen 1971; Brouw and Spoelstra 1976; Heiles 1996;
Beck 2001). This implies that the 
regular magnetic field is $4\pm 1$ $\mu$G at the
Solar radius. On the other hand, from Faraday rotation of pulsars and 
radio sources, Han et al.~(2006)
derived a regular field strength of $2.1\pm 0.3$ $\mu$G at the Sun position.
Possible explanations for the difference between the equipartition estimate
and the value inferred from pulsar data were discussed in Heiles (1996) and
Beck et al.~(2003). Since our aim is to place upper limits on the
magnetic effects, we will take the value derived from polarization
measurements, $\bar{B}_{\odot}\equiv \bar{B}_{\phi}(R_{\odot})\simeq 4$ $\mu$G, 
as a generous value.

The radial profile of $\bar{B}_{\phi}$ is not well constrained by
observations. In the inner Galaxy ($3\,{\rm kpc}<R<R_{\odot}$),
the ordered magnetic field gets stronger at smaller
Galactocentric radius, probably as $R^{-1}$ or $R^{-2}$
(Heiles 1996). Han et al.~(2006) used an exponential function
to fit the ordered magnetic field and found a scale radius
of $8.5\pm 4.7$ kpc in the radial interval between $3$ kpc and 
$R_{\odot}$. For the outer Galaxy ($R>R_{\odot}$), there is no
quantitative estimate of its exact $R$ dependence,
except that $\bar{B}_{\phi}\lesssim B_{\rm tot}$. 
As already said,
if the magnetic pitch angle is assumed to be constant with $R$, then
$\bar{B}_{\phi}\propto R^{-1}$. At $R<2R_{\odot}$, this radial
decay is consistent with WMAP7 22 GHz data
(Jansson \& Farrar 2012).

In order to illustrate how the results depend on the assumptions,
we will explore four different representative magnetic configurations
(see Figure \ref{fig:B_profiles}). 
In model A, we will assume that $\bar{B}_{\phi}$ declines
exponentially with $R$, in the outer Galaxy ($R>R_{\odot}$), with
the same scalelength as $B_{\rm tot}$: 
\begin{equation}
\bar{B}_{\phi}(R)=\bar{B}_{\odot}\exp\left(-\frac{R-R_{\odot}}{R_B}\right),
\label{eq:bphimodelA}
\end{equation}
with $\bar{B}_{\odot}=4$ $\mu$G and $R_{B}=12$ kpc.
As a consequence,
the ratio of the regular magnetic field
to the total magnetic field, $\eta$, is constant with radius
for $R>R_{\odot}$,
having a value of $\sim 0.6$. This is a well motivated possibility
because constant values for $\eta$ with galactocentric
distance have been derived
in external galaxies. For instance,
in the case of M31, Fletcher et al.~(2004)
derived $\eta\simeq 0.7$ in the radial range $8$ to $14$ kpc.
A rather constant value of $\eta$ within $R<6$ kpc was
found by Beck (2007) for the galaxy NGC 6946. 
In M33, Tabatabaei et al.~(2008) inferred a
value of $\eta\simeq 0.45$ independent of radius within $R<7$ kpc. 

In a second type of magnetic profiles (labeled as models B1 and B2), 
we will adopt the same
dependence for the azimuthal field of our Galaxy as in 
Ruiz-Granados et al.~(2012): 
\begin{equation}
\bar{B}_{\phi}(R)=\frac{(R_{l}+R_{\odot})
\bar{B}_{\odot}}{R_{l}+R},
\label{eq:mag_profile}
\end{equation}
where $\bar{B}_{\odot}=4$ $\mu$G and $R_{l}$ is, in principle, a free parameter.
To facilitate comparison with previous work, we will take $R_{l}=14$ kpc.
Models B1 and B2 have the same regular magnetic field as given
in Eq.~(\ref{eq:mag_profile}) but differ in the random magnetic component.
In model B1, we will assume that $\eta$ is constant with radius,
and has a value of $0.66$, rather similar to model A. 
Thus, $\left<b^{2}\right>=(\eta^{-2}-1) \bar{B}_{\phi}^{2}\simeq 1.3
\bar{B}_{\phi}^{2}$.
In model B2, the mean square turbulent field $\left<b^{2}\right>$
is obtained as the difference between
the total magnetic field as inferred from synchrotron emission
and the regular magnetic field, that is
$\left<b^{2}\right>=B_{\rm tot,fit}^{2}-\bar{B}_{\phi}^{2}$.
where $B_{\rm tot,fit}$ is given in Eq.~(\ref{eq:tot_fit})
with $B_{\rm tot,\odot}=6.9\,\mu$G.

\begin{figure*}
  \epsfig{file=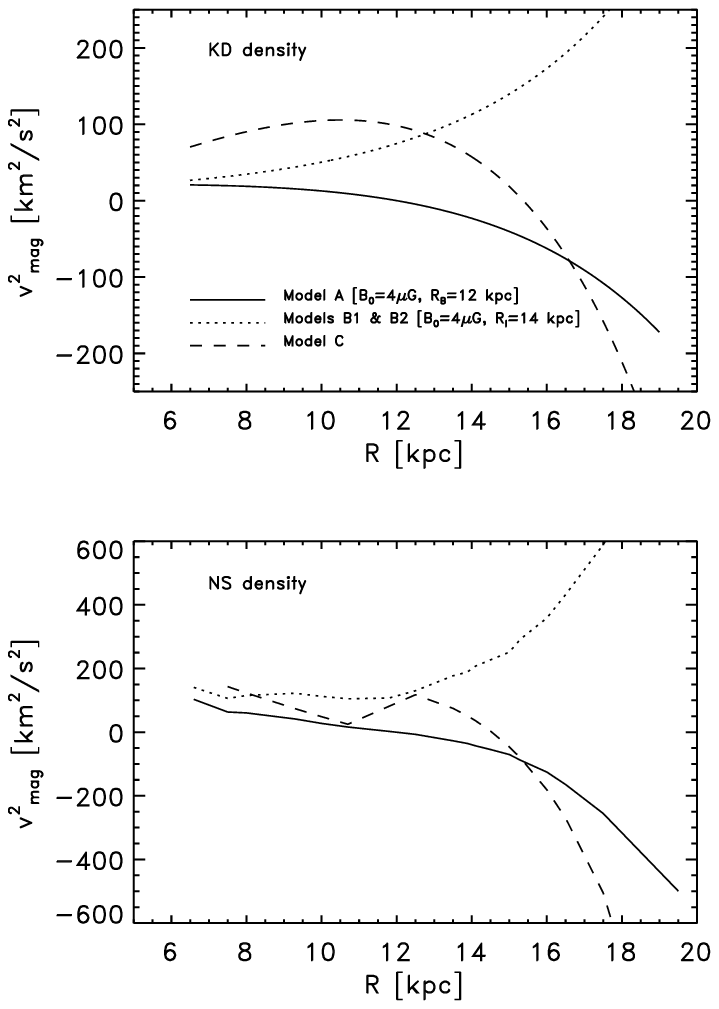,angle=0,width=12cm,height=16cm}
  \caption{
Magnetic contribution to the rotation curve
due to the azimuthal magnetic field, $v_{\rm mag}^{2}$,
as a function of radius, for models A, B1, B2 and C, using the
KD profile (upper panel) and the NS profile (lower panel). 	
The gas rotates at a speed of 
$v_{\phi}=(v_{c}^{2}+v_{P}^{2}+v_{\rm mag}^2)^{1/2}$.
  }
  \label{fig:comparison1}
\end{figure*}

Finally, we consider a fourth model (labeled as model C) in which 
we assume that there is equipartition between
the magnetic pressure in the random field and the dynamical pressure,
that is $P_{b}=P_{g}$ at any radius in the range $R_{\odot}<R<2R_{\odot}$. 
This is expected in turbulent discs where turbulent motions in the
gas tangle the magnetic field.
The equipartition condition determines $\left<b^{2}\right>$ as a function of
radius. Once $\left<b^{2}\right>$ is derived, the coherent magnetic 
field $\bar{B}_{\phi}$ is then
obtained as $\bar{B}_{\phi}^{2}=B_{\rm tot,fit}^{2}-\left<b^{2}\right>$.

Figure \ref{fig:B_profiles} shows the radial profiles for both the
strength of the azimuthal large-scale magnetic field and the 
strength of the small-scale random field $\left<b^{2}\right>^{1/2}$,
for the different models. We have assumed that $R_{\odot}=8.5$ kpc. 
Note that the magnetic profiles in model C depend on the adopted 
midplane H\,{\sc i} density; to make easier the discussion, 
we show the magnetic profiles
for our reference KD density profile. 
For comparison, we also plot
the total magnetic field as derived from synchrotron emission
(Beck et al.~1996; Beck 2001; Ferri\`ere 2001). 
By construction, models A, B2 and C fit very well the strength of the
total magnetic field at $R>7$ kpc.
Bearing in mind that the error in the determination of
$B_{\rm tot}$ from synchrotron emission data is  $30\%$, we can see that
models B1 are compatible with it at $R>7$ kpc. 

\begin{figure*}
  \epsfig{file=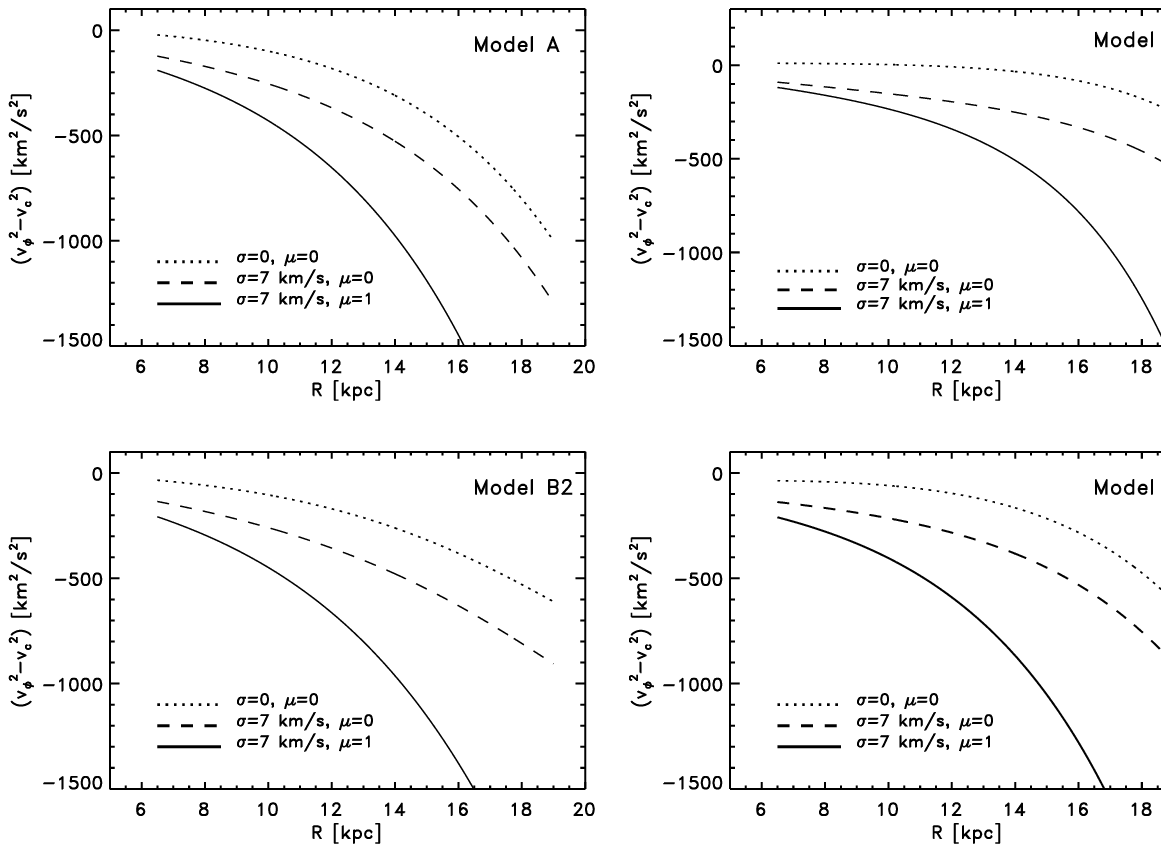,angle=0,width=16cm,height=15cm}
  \caption{
Difference between $v_{\phi}^{2}$ and $v_{c}^{2}$ for models A, B1, B2 
and C, and
different combinations of $\sigma$ and $\mu$. For the midplane
density we have used the reference KD profile. A negative (positive)
value means that the gas rotates slower (faster) than the gravitational
circular velocity. Cases with $\sigma=0$
correspond to ignoring the kinetic pressure of the gas, whereas
models with $\mu=0$ assume that the radial force by the
pressure of cosmic rays is null.  
  }
  \label{fig:all_terms}
\end{figure*}
\begin{figure*}
  \epsfig{file=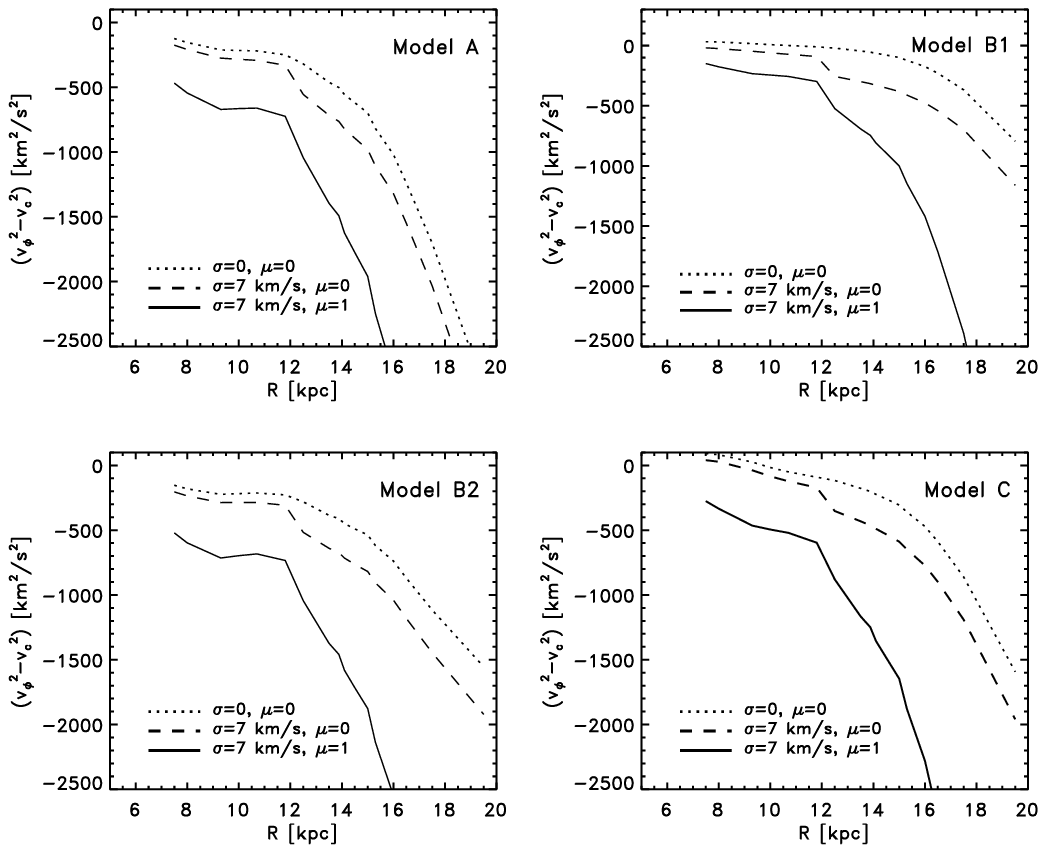,angle=0,width=16cm,height=15cm}
  \caption{
Same as Figure \ref{fig:all_terms} but for the NS gas profile.
For consistency, we use $R_{\odot}=8$ kpc here.
  }
  \label{fig:all_termsNS}
\end{figure*}

In model A, the azimuthal magnetic field varies between $4\,\mu$G
at $R_{\odot}$ to $2.0\,\mu$G at $2R_{\odot}$, and has  $\eta\simeq 0.6$
constant with radius. In model B1, $\bar{B}_{\phi}$ varies between $4\,\mu$G
and $2.9\,\mu$G, and $\eta$ is also constant with radius $\eta\simeq 0.66$.
Model B2 has the same azimuthal magnetic field as model B1 but $\eta$ 
increases radially from  $0.58$ at $R_{\odot}$ to $0.9$ at $2R_{\odot}$.
In model C plus the KD density profile, 
the azimuthal magnetic field varies between $4.6\,\mu$G at $R_{\odot}$
to $3.1\,\mu$G at $2R_{\odot}$, and $\eta$ increases from $0.68$ to $0.9$.
Finally, in model C plus the NS density profile,
$\bar{B}_{\phi}$ varies between $6.3\,\mu$G at $R_{\odot}$
to $3.4\,\mu$G at $2R_{\odot}$, and $\eta\geq 0.9$ at any radius
betweeen $R_{\odot}$ and $2R_{\odot}$.  Thus, in models B2 and C,
the magnetic field at $2R_{\odot}$ is dominated 
by the regular component. 
We should note here that the different radial profiles for $\bar{B}_{\phi}$
are realistic for a finite radial interval, $R_{\odot}<R<2R_{\odot}$,
but there is no reason to assume that they are equally realistic at large $R$
(e.g., S\'anchez-Salcedo \& Reyes-Ruiz 2004).

\section{Estimating the contributions to the rotation
curve}
\label{sec:results}
\subsection{Results}
We will start our discussion by considering the effect of the
azimuthal magnetic field in the rotation curve, $v_{\rm mag}^{2}$.
Figure \ref{fig:comparison1} shows $v_{\rm mag}^{2}$ as a function of $R$ for
the different models.  We see that 
the shape of $v_{\rm mag}^{2}$ as a function of $R$ depends critically 
on the adopted profile for $\bar{B}_{\phi}$.
In models A and C, $v_{\rm mag}^{2}$ is positive at small galactocentric
radii but turns out to negative values beyond a certain radius. 
Using Eqs.~(\ref{eq:vmag}) and (\ref{eq:bphimodelA}),
it is simple to show that, in model A, $v_{\rm mag}^{2}<0$ at $R>R_{B}$.
On the other hand, $v_{\rm mag}^{2}$ is positive at any radius in models 
B1 and B2.
The sign of $v_{\rm mag}^{2}$ at $2R_{\odot}$ is model-dependent 
and there is no clear
preference for a model with $v_{\rm mag}^{2}>0$ over another with 
$v_{\rm mag}^{2}<0$ at $2R_{\odot}$.  
Since, according to Equations (\ref{eq:vmag}) and (\ref{eq:vP}), 
the strength of the radial force (per unit of mass) 
by magnetic effects and cosmic rays is proportional to $\rho^{-1}$, 
the exact values for $v_{\rm mag}^{2}$ and $v_{P}^{2}$ 
depend on the H\,{\sc i} gas model.
If the NS profile is used, 
$v_{\rm mag}^{2}$ at $2R_{\odot}$
ranges between $-200$ km$^{2}$s$^{-2}$ to $400$ km$^{2}$s$^{-2}$ depending
on the model, whereas it varies between $-100$ km$^{2}$s$^{-2}$ to 
$220$ km$^{2}$s$^{-2}$ when the KD profile is used. 
Therefore, it has a minor 
effect on the rotational velocity of the gas;
the corresponding correction 
is  $\sim 0.5-1$ km s$^{-1}$ at $2R_{\odot}$.
At this galactocentric radius, the correction by $v_{\rm mag}^{2}$ is 
comparable to the correction by the kinetic
pressure of the gas. For instance, consider the H\,{\sc i} gas model
of KD. The pressure correction is
\begin{equation}
\frac{R}{\rho}\frac{dP_{g}}{dR}=-\frac{R}{R_{H}}\sigma^{2},
\label{eq:kinetic}
\end{equation}
which is $\simeq -260$ km$^{2}$s$^{-2}$ at $2R_{\odot}$ (using
$R_{H}=3.15$ kpc and $\sigma=7$ km s$^{-1}$, see \S \ref{sec:assump}
and \S \ref{sec:models}).
In the what follows, we discuss and quantify the relative importance
of $v_{P}^{2}$ as compared to $v_{\rm mag}^{2}$.

Figures \ref{fig:all_terms} and \ref{fig:all_termsNS}
show the contributions to the
tangential velocity of the gas, $v_{\phi}^{2}-v_{c}^{2}$, when
the magnetic pressure $P_{b}$ is included, 
for different combinations of $P_{g}$ and $P_{CR}$.
Obviously, the curves with $\sigma=0$ and $\mu=0$ correspond to
$P_{g}=P_{CR}=0$ [see Eqs.~(\ref{eq:kin_pressure}) and (\ref{eq:cr_pressure})].
On the other hand, curves with $\sigma=7$ km s$^{-1}$ and
$\mu=0$, include the kinetic pressure of the gas and the magnetic forces
(i.e., including both the azimuthal and the
small-scale components),
but not the pressure by cosmic rays.
The case $\mu=1$ describes a situation in which
the pressure by cosmic rays is in equipartition with
the magnetic pressure. For instance,  Ferri\`ere (2001) 
quotes a midplane value of $\mu=1.28$ in the vicinity of the Sun. 
In order to interpret correctly Figures \ref{fig:all_terms} and
\ref{fig:all_termsNS}, remind that 
when $v_{\phi}^{2}-v_{c}^{2}$ is negative, it means that
the MHD terms provide support to the disc and, hence,
the measured tangential velocity lags the circular velocity
of a test particle, i.e.~$v_{\phi}<v_{c}$.

\begin{figure}
  \epsfig{file=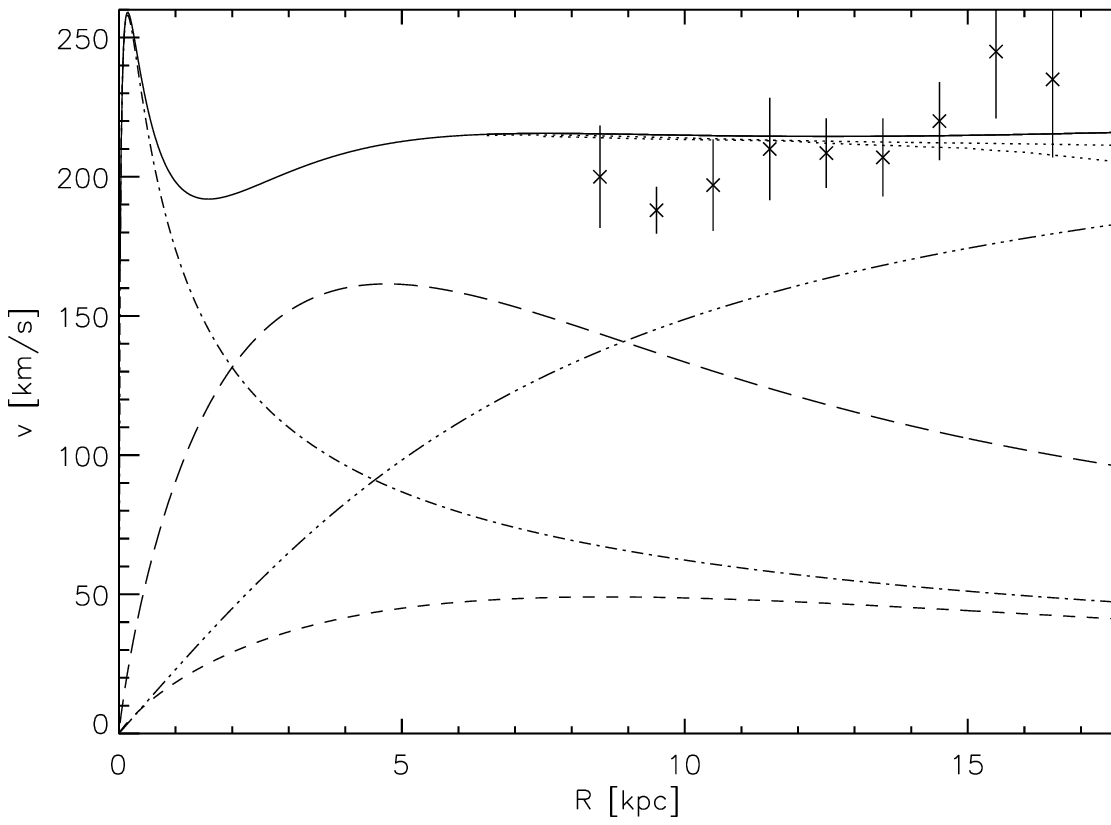,width=8cm,height=7cmm}
  \caption{
Total gravitational circular velocity $v_{c}$ (solid line) 
together with the tangential
velocity of the gas $v_{\phi}$ after including magnetic fields, gas pressure
 and cosmic rays in model A (dotted lines).  
The upper dotted line was derived using the H\,{\sc i} density profile
from KD, whereas the lower dotted line was
calculated using the profile derived in NS.
Symbols indicate the observed rotational velocity of H\,{\sc i}
gas taken from Ruiz-Granados et al.~(2012).
The contribution of the different mass
components to the rotation curve are also shown: 
bulge (dot-dashed line), stellar disc (long dashed line), gas 
(short dashed line) and dark halo (triple dot-dashed line). We see
that the difference between $v_{\phi}$ and $v_{c}$ is small as compared
to the observational uncertainties. }
  \label{fig:RC_18}
\end{figure}

In our models, the strength of the random field $\left<b^{2}\right>^{1/2}$
declines with radius and,
therefore, the magnetic pressure by this small-scale magnetic field
produces a force outwards, giving support to the disc (slower rotation).
In models A, B2 and C, this outward force is able to compensate
any pinching effect by the azimuthal magnetic field. 
Thus $v_{\phi}^{2}-v_{c}^{2}<0$ at any galactic radius $>7$ kpc.

As can be seen in Figure \ref{fig:all_terms} and
\ref{fig:all_termsNS}, the effect of magnetic fields in the
rotation curve is less important in model B1, even though it
presents the highest $B_{\rm tot}$-values at $2R_{\odot}$  (Figure
\ref{fig:B_profiles}). 
The reason is that the radial profiles of both $\bar{B}_{\phi}$ and
$\left<b^{2}\right>^{1/2}$ in model B1 are more shallow and, thus,
the confining effect of $\bar{B}_{\phi}$ is partly balanced by
the radial support of the random fields.
In model B1, the outward force by the radial gradient of $P_{b}$
is in balance with the inward pinching force created by $\bar{B}_{\phi}$
at a radius 
\begin{equation}
R_{B1}=\frac{\eta^{2}}{1-\eta^{2}}R_{l}.
\end{equation}
For $R_{l}=14$ kpc and $\eta=0.66$, this balance occurs at $R=10.8$ kpc.
In order to have $R_{B1}>20$ kpc in this kind of models, we need $\eta>0.76$. 
In model B2, the equality between the gradient of $P_{b}$ and
the inward force by $\bar{B}_{\phi}$ occurs in a more inner radius.
Model B2 illustrates the role of the magnetic pressure by the random field;
even if $\eta=0.9$ at $2R_{\odot}$, the contribution of
the random field to $v_{\phi}^{2}-v_{c}^2$, which is
$R\rho^{-1} dP_{b}/dR$, is three times
larger than $v_{\rm mag}^{2}$.

As already mentioned, the contribution of the kinetic 
pressure of the gas to the rotation velocity is comparable in 
magnitude to the magnetic terms and, therefore, it must be
included to estimate correctly the asymmetric drift $v_{\phi}^{2}-v_{c}^{2}$.
The same holds true for the cosmic ray pressure.
In our models, when all the terms are added, $v_{\phi}^{2}-v_{c}^{2}<0$
at any radius of interest. Hence the gas rotates at a speed less than $v_{c}$.
For the KD density profile with $\sigma=7$ km s$^{-1}$ and $\mu=1$,
we infer that $v_{P}^{2}+v_{\rm mag}^{2}$ at $2R_{\odot}$
ranges between $-1000$ to $-1850$ km$^{2}$s$^{-2}$, 
and between $-1500$ to $-2600$ km$^{2}$s$^{-2}$
if the NS density is used.
Figure \ref{fig:RC_18} shows the gravitational circular velocity 
for a typical mass model of the Galaxy,
together with the tangential velocity of the gas in model A, when all the
correction terms in the asymmetric drift are included.
We see that the effect is very small as compared to the intrinsic
uncertainties in the determination of the H\,{\sc i} rotational velocity; 
at $2R_{\odot}$ the gas is expected
to rotate about $\lesssim 4-8$ km s$^{-1}$ slower than the corresponding
gravitational circular velocity $v_{c}$ at this radius. We find that
for realistic models,
the magnetic fields and cosmic rays cannot rise the rotation curve
at galactocentric distances of $\sim 2R_{\odot}$,
contrary to the suggestion by Ruiz-Granados et al.~(2012).  

\begin{figure*}
  \epsfig{file=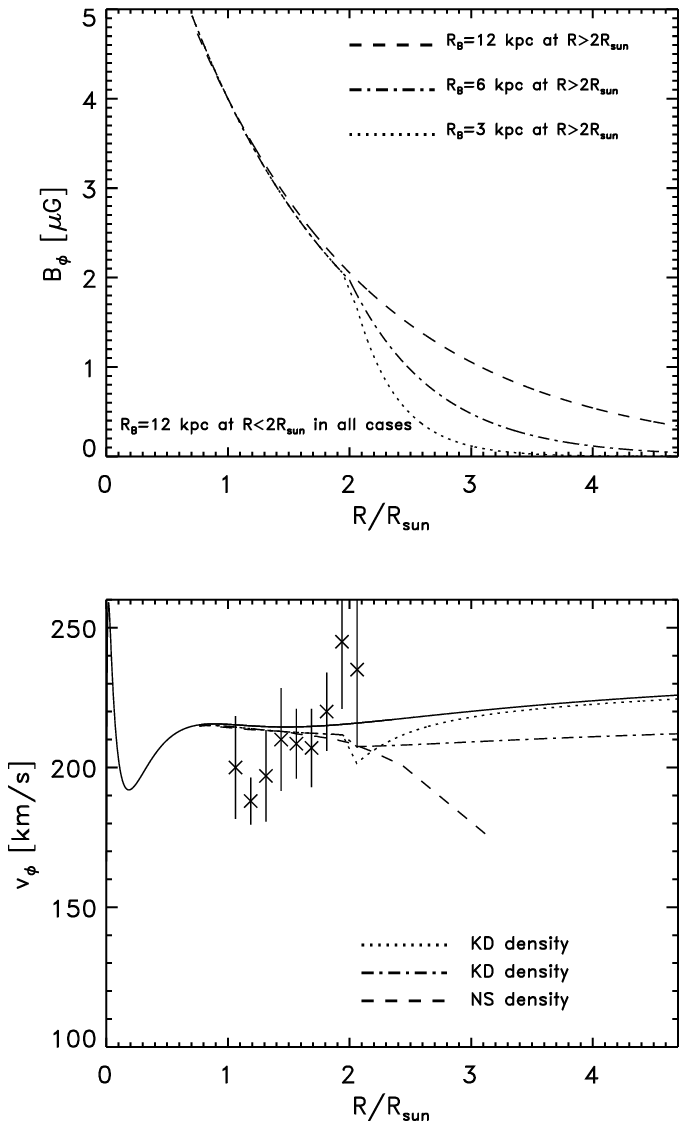,angle=0,width=13cm,height=18cm}
\vskip 0.4cm
  \caption{
Azimuthal magnetic field and the corresponding effect on the gas
rotation velocity for a double piece-wise exponential magnetic
field, with $\eta=0.6$, $\sigma=7$ km s$^{-1}$ and $\mu=1$, all
constant with radius.  
In all the models, the magnetic field declines exponentially 
with a scalelength of $12$ kpc up to $2R_{\odot}$.
Beyond $2R_{\odot}$ kpc, the scalelength of the magnetic field is $3$ kpc
(dotted line), $6$ kpc (dot-dashed line) and $12$ kpc (dashed line).
The mass model is the same as in Figure \ref{fig:RC_18}. 
Symbols indicate the observed rotational velocity of H\,{\sc i}
gas taken from Ruiz-Granados et al. (2012). Note that
the $y$-axis ranges from $100$ to $260$ km s$^{-1}$.}
  \label{fig:RC_40}
\end{figure*}

It is interesting to note that Figures \ref{fig:all_terms} and 
\ref{fig:all_termsNS} show that the correction to the
rotation curve by magnetic fields and cosmic rays increases steeply with 
galactocentric radius. In particular, if we extrapolate model A 
to larger galactocentric distances and use the NS gas profile, we
would find 
that the gas at $3R_{\odot}$ would rotate $\sim 40$ km s$^{-1}$
slower than a test particle on a circular orbit 
(see Figure \ref{fig:RC_40}).
However,  the magnetic profiles used in our models are based
on synchrotron observations at $R<2R_{\odot}$. Thus, there is no
reason to assume that these profiles are valid at any $R$.
In order to explore how $v_{\phi}$ depends on the adopted magnetic
profile, 
Figure \ref{fig:RC_40} shows $v_{\phi}$ in models with $\eta=0.6$
(constant with radius),
$\sigma=7$ km s$^{-1}$ and $\mu=1$, where the azimuthal
magnetic field is described by a double piece-wise exponential 
profile; $R_{B}=12$ kpc at $R<2R_{\odot}$ kpc (as in model A)
but having a steeper radial decline beyond $2R_{\odot}$. 
We see that when $R_{B}\simeq 6$ kpc in the outer disc ($R>2R_{\odot}$),
the MHD terms produce a shift in the azimuthal velocity
of $\sim 15$ km s$^{-1}$ (the KD density
profile was used). On the other hand,
if the outer magnetic field declines with a scalelength of $3$ kpc, the radial
support by cosmic rays and magnetic fields leads to   
$v_{c}-v_{\phi}>5$ km s$^{-1}$ only between $2R_{\odot}$ and $2.6R_{\odot}$.
Beyond $2R_{\odot}$, empirical determinations of the strength/topology of
magnetic fields and the H\,{\sc i} rotation curve are challenging and, 
hence, there is no means of testing the effect of magnetic fields
on the H\,{\sc i} azimuthal velocity.  
Still, one has to consider the vertical confinement of the magnetic
fields. Consider model A with a {\it single} exponential scalelength
of $R_{B}=12$ kpc. At $3R_{\odot}$, the cosmic ray plus magnetic pressure in 
the midplane is $\sim 0.2\times 10^{-12}$ dyn cm$^{-2}$. Since
the weight of neutral gas cannot account for this large pressure,
an additional coronal component should be invoked to provide
vertical support. Following the same analysis as Cox (2005) did 
at the solar neighbourhood, we find that a component with 
density ($0.003$cm$^{-3})\exp(-|z|/z_{g})$, with $z_{g}\sim 10-12$ kpc,
and temperature
of $\sim 3\times 10^{5}$ K (at $3R_{\odot}$) could confine the cosmic rays 
and magnetic field at $3R_{\odot}$. This coronal layer would
probably produce excessive X-ray emission and the total column
density of OVI would be orders of magnitude more than is
observed looking out of the galactic plane (Cox 2005).
It is more simple to assume that beyond the stellar truncation
radius, stellar formation is almost nonexistent, so energy
in cosmic rays and magnetic fields decays faster with $R$ because
the energy input into these components from stellar processes
is likely to be less important (e.g., Olling \& Merrifield 2000).

\subsection{Other magnetic models}
We have shown that in all our models, the magnetic fields provide 
support to the disc when the magnetic pressure by the random field component 
is included (at least beyond a galactocentric distance of $10.8$ kpc). 
If we had adopted $\bar{B}_{\phi} =2 \mu$G 
instead of $4\mu$G, the magnetic pressure by the random field
would increase by a factor of $1.6$ 
(in order to account for the synchrotron radio emission),
providing more radial support to the disc. 
Consider, for instance, model A with the KD profile for 
$\sigma=7$ km s$^{-1}$ and $\mu=1$.
At $2R_{\odot}$, $v_{P}^{2}+v_{\rm mag}^{2}$ would change
from $-1850$ km$^{-2}$ s$^{-2}$ for $\bar{B}_{\odot}=4\mu$G, to
$-2250$ km$^{2}$ s$^{-2}$ if a lower value $\bar{B}_{\odot}=2\mu$G
is adopted.

In order for the gas to rotate faster than a test particle, 
we need $v_{P}^{2}+v_{\rm mag}^{2}>0$.
This is possible only in models
in which the radial gradients in the pressure created by the random 
magnetic field and cosmic rays are small.
To explore this possibility, we have treated
$R_{l}$ in model B1 as a free parameter and allowed to vary
until $v_{P}^{2}+v_{\rm mag}^{2}$ reaches a maximum at $2R_{\odot}$.
We found that the maximum occurs when the radial scale length is very large
so that $\bar{B}_{\phi}$,
$P_{b}$ and $P_{CR}$ are constant with galactic radius.
In such a configuration, $B_{\rm tot}\simeq 6$ $\mu$G, constant
in the range $R_{\odot}<R<2R_{\odot}$, which
is inconsistent with the exponential radial
decline observed in the synchrotron emission.
In addition, a constant pressure of cosmic rays can hardly
be maintained if the sources of cosmic rays are related to star-forming
regions in the inner galaxy and they diffuse outwards losing energy
(e.g., Everett et al.~2010).
Still, it is worthwhile to consider what happens in this
unlikely situation. Since the pressure gradients by the turbulent
magnetic field and cosmic rays are null, only  
the kinetic pressure and the $v_{\rm mag}^{2}$-term play a role.
Combining both contributions and assuming that $\rho$ declines
exponentially with $R$, we find 
\begin{equation}
v_{P}^{2}+v_{\rm mag}^{2}=\frac{\bar{B}_{\odot}^{2}}{4\pi\rho_{\odot}}
\exp\left(\frac{R-R_{\odot}}{R_{H}}\right)-\frac{R}{R_{H}}\sigma^{2}.
\end{equation}
For $\bar{B}_{\odot}=4$ $\mu$G, $R_{H}=3.15$ kpc and $\sigma=7$ km s$^{-1}$,
we infer $v_{P}^{2}+v_{\rm mag}^{2}=637$ km$^{2}$s$^{-2}$.
The correction to the tangential velocity of the gas is $1.5$ km s$^{-1}$
if $v_{c}=220$ km s$^{-1}$ or $1.8$ km s$^{-1}$ if 
$v_{c}=175$ km s$^{-1}$. 
We conclude that, in axisymmetric models,  
the observed tangential velocity is not expected to differ significantly
from the true gravitational circular velocity at the interval
$R_{\odot}<R<2R_{\odot}$.

\subsection{Comparison with previous works}
Ruiz-Granados et al.~(2012) 
claim that a significant improvement of the fit to the rotation
curve of the Milky Way is obtained
when magnetic fields are considered.  In their modelling, they
only include the azimuthal component and ignore any contribution
from the turbulent component of the magnetic field, kinetic pressure
or cosmic rays. 
They use the same expression for the azimuthal magnetic field
as that given in Eq.~(\ref{eq:mag_profile})
and find the values of $R_{l}$ that provide the best fit to the shape of 
the rotation curve of the Milky Way.
They find $R_{l}=14.2^{+2.04}_{-4.17}$ kpc in a mass model with
a pseudo-isothermal dark halo and  
$R_{l}=16.5\pm 1.1$ kpc if there is no dark matter at all within
a sphere of radius $2R_{\odot}$.

\begin{figure}
  \epsfig{file=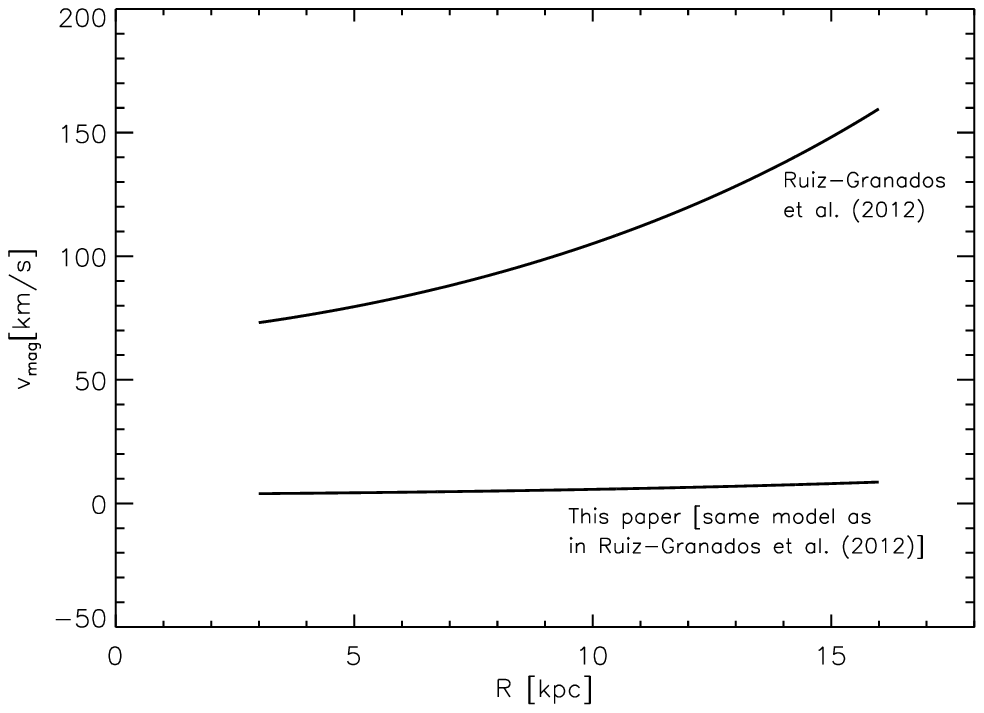,angle=0,width=8cm,height=7cm}
\vskip 0.4cm
  \caption{Magnetic contribution to the rotation curve due to the
azimuthal magnetic field, $v_{\rm mag}$, as a function of radius,
for $R_{\odot}=8$ kpc, $\bar{B}_{\phi}=3\,\mu$G, $R_{H}=4$ kpc 
and $R_{l}=14.2$ kpc.
These parameters correspond to the best fit model denoted by
ISO$+$MAG in Ruiz-Granados et al.~(2012). 
The values of $v_{\rm mag}$ were
overestimated by a factor of $18$ in Ruiz-Granados et al.~(2012).} 
\label{fig:comparison}
\end{figure}

The exact values for $\bar{B}_{\odot}$, $\rho_{\odot}$, $R_{\odot}$
and $R_{H}$ in Ruiz-Granados et al.~(2012) differ from those adopted 
in this paper, but only slightly.
They used $\bar{B}_{\odot}=3$ $\mu$G, $R_{\odot}=8$ kpc, $R_{H}=4$ kpc,
a column density of gas at the Sun position of $10M_{\odot}$pc$^{-2}$,
and a constant vertical scale height of $0.2$ kpc across the disc.
Figure \ref{fig:comparison} shows 
$v_{\rm mag}$, as a function of $R$, for $R_{l}=14.2$ kpc and
the abovementioned values for $\bar{B}_{\odot}$, $R_{\odot}$, $R_{H}$ 
and $\rho_{\odot}$.
A comparison with the corresponding curve reported in figure 2 of
Ruiz-Granados et al.~(2012) dictates that $v_{\rm mag}$ was overestimated
by a factor of $18$.
For a model with $R_{l}=16.5$ kpc, we obtain  
$v_{\rm mag}=9.5$ km s$^{-1}$ at $2R_{\odot}$, which is too small
to math the circular velocity
without any dark matter (a value of $v_{\rm mag}\sim 180$ km s$^{-1}$
is required to do so).
Even if we neglect the radial support by the cosmic-ray pressure and
by the turbulent magnetic field,
and only include the kinetic pressure
of the gas, we infer $v_{P}^2+v_{\rm mag}^{2}\simeq -100$ km$^{2}$s$^{-2}$ 
at $2R_{\odot}$ in this model ($R_{l}=16.5$ kpc). That is unable to
provide the desired effect in the rotation curve.

\section{Conclusions}
\label{sec:conclusions}
How magnetic effects alter the overall rotation curve of gas in galaxies
is a reoccurring theme in the literature.
In a recent paper, Ruiz-Granados et al.~(2012) 
claim that magnetic field forces provide the simplest way to explain
the peculiar rising-up of the rotation curve in our Galaxy.
Our Galaxy offers 
a unique opportunity for studying the three-dimensional
distribution of neutral gas and magnetic fields in a detail
unobtainable in external galaxies.
We have explored a range of plausible models to quantify
the contribution to the radial support by kinetic gas pressure,
magnetic fields, and cosmic rays. We restrict ourselves
to the interval $R_{\odot}<R<2R_{\odot}$, because beyond $2R_{\odot}$
there are no determinations of the strength of the magnetic field. 
We have shown that, even adopting magnetic
field configurations with a regular field of $\sim 3\mu$G at
$2R_{\odot}$, the rotation curve of our Galaxy is not appreciably 
altered by magnetic effects. 
Turbulent motions, cosmic rays and the random small-scale component of the 
galactic magnetic fields act as pressure, giving support to the disc and,
therefore, leading to a rotation a few km s$^{-1}$
slower than the gravitational circular speed.  
Given the large uncertainties 
in the rotation speed of the outer parts of
the Galaxy, of $\pm 25$ km s$^{-1}$, they can be safely ignored
at least within $R<2R_{\odot}$.

\section*{acknowledgements}
We would like to thank A.~Fletcher, J.~Franco, J.~A.~Garc\'{\i}a-Barreto
and P.~Kalberla for fruitful discussions.
We acknowledge financial support from CONACyT project 165584
and PAPIIT project IN106212.


\begin{thebibliography}{}
\bibitem[Battaner et al.(1992)]{bat92}
Battaner, E., Garrido, J. L., Membrado, M., Florido, E. 1992, Nature, 360, 652
\bibitem[Beck(2001)]{bec01}
Beck, R. 2001, Space Science Reviews, 99, 243
\bibitem[Beck(2007)]{bec07}
Beck, R. 2007, A\&A, 470, 539
\bibitem[Beck et al.(1996)]{bec06}
Beck, R., Brandenburg, A., Moss, D., Shukurov, A., \&  Sokoloff, D.
1996, ARA\&A, 34, 155
\bibitem[Beck et al.(2003)]{bec03}
Beck, R., Shukurov, A., Sokoloff, D., \& Wielebinski, R.
2003, A\&A, 411, 99
\bibitem[Berkhuijsen(1971)]{ber71}
Berkhuijsen, E. M. 1971, A\&A, 14, 359
\bibitem[Blitz \& Spergel(1991)]{bli91}
Blitz, L., \& Spergel, D. N. 1991, ApJ, 370, 205
\bibitem[Brouw \& Spoelstra(1976)]{bro76}
Brouw, W. N., \& Spoelstra, T. A. Th. 1976, A\&AS, 26, 129
\bibitem[Burton et al.(1992)]{bur92}
Burton, W. B. 1992, Distribution and Observational Properties of the
ISM, in The galactic interstellar
medium (Springer-Verlag), Saas-Fee Advanced Course 21,
ed.~D.~Pfenniger, P.~Bartholdi, Springer-Verlag, 1
\bibitem[Cox(2005)]{cox05}
Cox, D. P. 2005, ARA\&A, 43, 337
\bibitem[Cuddeford \& Binney(1993)]{cud93}
Cuddeford, P., Binney, J. J. 1993, Nature, 365, 20
\bibitem[Dalcanton \& Stilp(2010)]{dal10}
Dalcanton, J. J., \& Stilp, A. M. 2010, ApJ, 721, 547
\bibitem[de Blok \& Bosma(2002)]{blo02}
de Blok, W. J. G., \& Bosma, A. 2002, A\&A, 385, 816 
\bibitem[Dib et al.(2006)]{dib06}
Dib, S., Bell, E., \& Burkert, A. 2006, ApJ, 638, 797
\bibitem[Everett et al.(2010)]{eve10}
Everett, J. E., Schiller, Q. G., \& Zweibel, E. G. 2010, ApJ, 711, 13
\bibitem[Ferri\`ere(2001)]{fer01}
Ferri\`ere, K. M. 2001, Rev.~Modern Physics, 73, 1031
\bibitem[Fletcher et al.(2004)]{fle04}
Fletcher, A., Berkhuijsen, E. M., Beck, R., \& Shukurov, A. 2004, A\&A,
414, 53
\bibitem[Han et al.(2006)]{han06}
Han, J. L., Manchester, R. N., Lyne, A. G., Qiao, G. J., \& van Straten, W.
2006, ApJ, 642, 868
\bibitem[Heiles(1996)]{hei96} 
Heiles, C. 1996, in Polarimetry of the interstellar medium, 
W. Roberge and D. Whittet eds., ASP Conf. Ser., San Francisco 97,  457
\bibitem[Ja{\l}ocha et al.(2012a)]{jal12a}
Ja{\l}ocha, J., Bratek, {\L}., Pekala, J., \& Kutschera, M. 2012a, MNRAS, 421, 2155
\bibitem[Ja{\l}ocha et al.(2012b)]{jal12b}
Ja{\l}ocha, J., Bratek, {\L}., Pekala, J., \& Kutschera, M. 2012b, 
MNRAS, 427, 393
\bibitem[Jansson \& Farrar(2012)]{jan12}
Jansson, R., \& Farrar, G. R. 2012, ApJ, 761, L11
\bibitem[Kalberla et al.(2008)]{kal08}
Kalberla, P. M. W., \& Dedes, L. 2008, A\&A, 487, 951
\bibitem[Kalberla et al.(2007)]{kal07}
Kalberla, P. M. W., Dedes, L., Kerp, J., \& Haud, U. 2007, A\&A, 469, 511
\bibitem[Lou \& Fan(1998)]{lou98}
Lou, Y.-Q., \&  Fan, Z. 1998, MNRAS, 493, 102
\bibitem[Nakanishi \& Sofue(2003)]{nak03}
Nakanishi, H., \& Sofue, Y. 2003, PASJ, 55, 191
\bibitem[Nelson(1988)]{nel88}
Nelson, A. H. 1988, MNRAS 233, 115
\bibitem[Olling \& Merrifield(2000)]{oll00}
Olling, R. P., \& Merrifield, M. R. 2000, MNRAS, 311, 361
\bibitem[Parker(1966)]{par66}
Parker, E.N. 1966, ApJ, 145, 811
\bibitem[Ruiz-Granados et al.(2010)]{rui10}
Ruiz-Granados, B., Rubi\~no-Mart\'{\i}n, J. A., Florido, E.,
\& Battaner, E. 2010, ApJ, 723, L44
\bibitem[Ruiz-Granados et al.(2012)]{rui12}
Ruiz-Granados, B., Battaner, E., Calvo, J., Florido, E.,
Rubi\~no-Mart\'{\i}n, J. A. 2012, ApJ, 755, L23
\bibitem[S\'{a}nchez-Salcedo(1997a)]{san97a}
S\'{a}nchez-Salcedo, F. J. 1997a, MNRAS, 289, 863
\bibitem[S\'{a}nchez-Salcedo(1997b)]{san97b}
S\'{a}nchez-Salcedo, F. J. 1997b, Ap\&SS, 249, 223 
\bibitem[S\'{a}nchez-Salcedo \& Reyes-Ruiz(2004)]{san04}
S\'{a}nchez-Salcedo, F. J., \& Reyes-Ruiz, M. 2004, ApJ, 607, 247 
\bibitem[Spitzer(1978)]{spi78}
Spitzer, L. 1978, Physical processes in the interstellar medium,
John Wiley \& Sons (New York)
\bibitem[Strong et al.(2000)]{str00}
Strong, A. W., Moskalenko, I. V., \& Reimer, O. 2000, ApJ, 537, 763
\bibitem[Tabatabaei et al.(2008)]{tab08}
Tabatabaei, F. S., Krause, M., Fletcher, A., \& Beck, R. 2008,
A\&A, 490, 1005
\bibitem[Tsiklauri(2011)]{tsi11}
Tsiklauri, D. 2011, Ap\&SS, 334, 165
\bibitem[Vall\'ee(1994)]{val94}
Vall\'ee, J. P. 1994, ApJ, 437, 179

\end{thebibliography}
\end{document}